\documentclass[manuscript,screen]{acmart}
\renewcommand\footnotetextcopyrightpermission[1]{} 
\AtBeginDocument{%
  }

\setcopyright{none}

\acmSubmissionID{1000}

\usepackage{verbatim}
\usepackage{amsmath}
\usepackage{hyperref}
\usepackage{booktabs}
\usepackage{enumitem}
\usepackage{soul}
\usepackage{hyperref}

\newcommand{\alex}[1]{{\color{teal}[Alex: #1]}}

\DeclareMathOperator*{\argmax}{arg\,max}

\newcommand{\mstate}{\theta}
\newcommand{\statespace}{\Theta}

\newcommand{\joint}{\pi}
\newcommand{\jointapprox}{\hat{\pi}}

\newcommand{\dist}{p}

\newcommand{\posterior}{q}

\newcommand{\distover}[1]{\Delta(#1)}

\newcommand{\action}{a}
\newcommand{\actionspace}{\mathcal{A}}

\newcommand{\signal}{v}
\newcommand{\signalspace}{\mathcal{V}}

\newcommand{\data}{\textbf{x}}
\newcommand{\dataRV}{X}
\newcommand{\dataspace}{\mathcal{X}}

\newcommand{\score}{S}
\newcommand{\proper}{\hat{\score}}

\newcommand{\reals}{{\mathbb R}}

\usepackage{ifthen}

\newcommand{\prob}[2][]{\text{\bf Pr}\ifthenelse{\not\equal{}{#1}}{_{#1}}{}\![{\def\givenn{\middle|}#2}]}
\newcommand{\expect}[2][]{\text{\bf E}\ifthenelse{\not\equal{}{#1}}{_{#1}}{}\![{\def\givenn{\middle|}#2}]}

\begin{document}

\title{Decision Theoretic Foundations for Experiments Evaluating Human Decisions}

\author{Jessica Hullman}
\email{jhullman@northwestern.edu}
\orcid{}
\affiliation{%
  \institution{Northwestern University}
  \city{Evanston}
  \state{Illinois}
  \country{USA}
}

\author{Alex Kale}
\email{kalea@uchicago.edu}
\orcid{}
\affiliation{%
  \institution{University of Chicago}
  \city{Chicago}
  \state{Illinois}
  \country{USA}
  }

\author{Jason Hartline}
\email{hartline@eecs.northwestern.edu}
\orcid{}
\affiliation{%
  \institution{Northwestern University}
  \city{Evanston}
  \state{Illinois}
  \country{USA}
}

\renewcommand{\shortauthors}{Hullman et al.}

\begin{abstract}
Decision-making with information displays is a key focus of research in areas like human-AI collaboration and data visualization. However, what constitutes a decision problem, and what is required for an experiment to conclude that decisions are flawed, remain imprecise. We present a widely applicable definition of a decision problem synthesized from statistical decision theory and information economics. We claim that to attribute loss in human performance to bias, an experiment must provide the information that a rational agent would need to identify the normative decision. We evaluate whether recent empirical research on AI-assisted decisions achieves this standard. We find that only 10 (26\%) of 39 studies that claim to identify biased behavior presented participants with sufficient information to make this claim in at least one treatment condition. We motivate the value of studying well-defined decision problems by describing a characterization of performance losses they allow to be conceived. 
\end{abstract}

\maketitle

\section{Introduction}\label{sec:intro}
\noindent Identifying effective information displays to support data-driven inference and decision-making is a common goal across a number of domains, including research related to human-computer interaction (HCI). For example, visualization researchers emphasize assisting decision-making as a vital goal of data visualization~\cite{dimara2021critical}.
Similarly, in human-centered artificial intelligence (AI), a growing body of empirical studies on human decision behavior has been called 
``necessary to \textit{evaluate} the effectiveness of AI technologies in assisting decision making, but also to form a \textit{foundational understanding} of how people interact with AI to make decisions''~\cite{lai2023towards}. 
Human decisions from predictions and other information displays are often evaluated in studies with the goal of identifying forms of bias: ways in which study participants' behavior deviates from expectations of good decision-making. 
However, within HCI and related fields there is little consensus on what a decision is, and how decisions are a distinct form of human behavior compared to other tasks.

For example,  
colloquial definitions describe a decision task as a choice between alternatives, or a choice where the stakes are high (such as deciding which of multiple hurricane forecasts is more accurate~\cite{padilla2015influence}, or whether to screen child welfare cases~\cite{zytek2021sibyl}). Neither isolates a clear set of assumptions that must hold for a judgment to qualify as a decision.
Other accounts cast decision-making as a higher-level task akin to sensemaking~\cite{dimara2021critical}.
For some, the point of studying human decisions is to compare them to a definition of idealized behavior. Sometimes 
researchers define this standard in a way that is common across studies (e.g., a better forecast comes closer to the true probability of the event~\cite{green2019disparate,fogliato2021impact}). Other times, however, different definitions of the same concept are used: e.g., appropriate reliance as agreeing with an AI when it is correct and disagreeing when it's not~\cite{bansal2021does,buccinca2021trust}, versus providing a higher subjective rating of trust when the model is correct~\cite{bussone2015role}, versus choosing the better of the human versus the AI prediction to bet on for a future task~\cite{dietvorst2015algorithm}.
Sometimes the point is to study seemingly ``subjective decisions,'' like how people react to music recommendations~\cite{kulesza2012tell}, or decide whether a house seems to match its valuation~\cite{prabhudesai2023understanding}, where it can seem unclear whether normative interpretations of task performance 
can be applied.

Not only is a more coherent definition of decision needed, as recent survey papers recommend~\cite{dimara2021critical,lai2021towards}, but such a definition must clearly define requirements for \textit{evaluating} decisions by identifying forms of bias or performance loss in human decision-making. 
In particular, we are concerned with the minimum requirements for a controlled study of human behavior to support claims that the human decision-making is flawed or non-ideal in some way. While it is generally recognized that such evaluative studies require ground truth outcomes to be identifiable for decision instances provided to participants, methodological work on experimental design in HCI and beyond lacks much formal guidance on what constitutes a valid research design for drawing normative conclusions about human decisions.


This paper contributes a definition of a decision problem and associated guidelines for experiment design that apply to any study intended to evaluate human decision-making under the provision of information. We show how a widely applicable framework for studying decision-making can be synthesized from several bodies of work outside HCI. First, foundations of statistical decision theory and expected utility theory~\cite{savage1972foundations,steele2015decision,von2007theory} rigorously define what it means to make good decisions under uncertainty. 
Second, information economics formalizes information structures that capture the relevant information in a set of signals for a decision problem~
\cite{bergemann2019information}. 
We argue that for a decision problem to be well-defined, it must be possible to define a normative (best response) decision to the information provided to the study participants. In other words, whether the results of a study can be used to argue that humans are biased depends on the answer to the question: Could a rational agent who perfectly perceived the information have used it to identify a best response? 
Importantly, this best response is calculated relative to the information available to the participant. This makes it distinct from scoring responses directly against the ground truth outcome. As an example of the difference, imagine a study where participants are shown weather forecasts, and asked to make decisions about whether to bring an umbrella or not. Rather than concluding that a participant made the wrong decision to bring the umbrella because a simulated outcome was no rain, they should be scored relative to whether the choice they made maximized their expected utility according to the scoring function the researcher uses to tabulate results. Additionally, this scoring function should have been communicated to them if the researchers want to draw conclusions about how good their decisions were.

We motivate this criterion by showing how performance loss can only be conceived of when a study participant could in theory identify how to optimize their behavior from the information they are provided. If, for example, the researchers had not communicated to participants of the weather forecast study sufficient information about how their responses would be judged, then we should expect different participants to make different guesses about what their goals should be. Consequently, their responses are incomparable, and it is left ambiguous what decision problem they thought they were solving. On the other hand, if they were provided sufficient information on the scoring process and provided forecast information, it is in theory possible to distinguish between a limited set of errors that explain observed performance loss, such as not extracting all of the decision-relevant information from the display, or not being able to identify the best response given the information they did extract.

To assess the extent to which recent empirical evaluations in HCI achieve this ideal, we apply our definition to 46 recent studies on AI-assisted decision-making, finding that while 39 of the studies draw conclusions about shortcomings in human decisions with prediction displays, only 10 (26\%) of these studies provided participants with a well-defined decision problem in at least one treatment condition.

The benefits of adopting a more rigorous standard based in established theoretical foundations for decision-making~\cite{savage1972foundations,bergemann2019information} for empirical decision research in HCI, human-centered AI/ML, visualization, and related fields are twofold.
First, it stands to improve the validity of interpretations of individual studies. The variation induced when a decision problem is not clearly reported nor communicated to participants confounds our ability to interpret the results, in that it cannot be separated from variation induced by true differences in how people do a task under different conditions. 
Adopting a standard where the normative decision arises from a well-defined process puts researchers in a better position to point to possible sources of performance loss,  and raises awareness of the limitations on what can be concluded from study results when a decision problem is under-communicated. 
Second, it paves the way for the development of theory. Only when a decision problem is well-defined can human responses to that problem be interpreted relative to the results of other studies (e.g., that explore variations on the problem). This provides HCI researchers an ability to make meaningful contrasts to drive theory development. Otherwise, any apparent differences in the quality of behavior may reflect noise induced by presenting participants with an underspecified problem.

The paper is organized as follows. First, we define a decision problem, normative decision, and sources of performance loss for human decision-makers. Next, we provide study design recommendations for applying this framework in practice. We present two example study designs in detail, and describe how they could be improved to align with the framework we present. We apply the framework to a sample of studies on AI-assisted decisions, illustrating the gap between the ideal it presents and current practice. We conclude with a discussion of limitations and future work.

\section{Defining a Decision Problem}\label{sec:def}
We define a decision problem and the normative decision given such a problem and enumerate possible sources of loss. 
Our definitions are intended for controlled evaluation of human behavior, a.k.a. \textit{normative} decision research. Behavioral data (which may be produced by humans or by simulation) is collected under controlled conditions, with the goal of learning about behavior induced by the provision of information. 
Such studies are frequently used to describe the quality of human performance in some situation (e.g., how well people make decisions from displays in strategic settings~\cite{zhang2023designing}), to rank different assistive elements by human performance (e.g, different visualizations~\cite{correll2018value} or AI explanation strategies~\cite{liu2021understanding}), or to test a hypothesis about how humans make decisions or what will help them do better (e.g., cognitive forcing functions will improve AI-assisted decisions~\cite{buccinca2021trust}).
Critically, such evaluative studies require that ground truth can be identified for any hypothetical states of the world against which study participants' decisions are evaluated.


\subsection{Decision Problem}
\label{sec:dec_prob}
Our definition of a \textbf{decision problem} for controlled study starts by assuming a scenario
in which there is some state of the world $\mstate$ which is uncertain at the time of the decision (\textbf{uncertain state}). The state is drawn from a \textbf{state space} $\statespace$ that describes the set of finite, mutually exclusive values that it can take. 

The \textbf{data-generating model} is a distribution over values of the uncertain state and \textit{signals} that provide information to the agent about the state. The data-generating model associates the state with the information that the agent is presented with (the \textbf{signal}) in order to inform their decision; for example a model prediction or visualization of relevant data. 
Since the signal is correlated with the state, if the participant understands it in the context of the decision problem, they can improve their performance compared to a case where they do not have access to the signal. 
More precisely, the data-generating model defines a joint distribution (or information structure) $\joint \in \distover{\signalspace \times \statespace}$ over signals $\signal \in \signalspace$ and states $\mstate \in \statespace$.
This joint distribution assigns to each realization $(\signal,\mstate) \in \signalspace \times \statespace$ a
probability, denoted $\joint(\signal,\mstate)$.

If we ignore the signal $\signal$, $\joint$ gives rise to a probability distribution over values of the uncertain state, which we denote as the prior distribution $\dist \in \distover{\statespace}$. The prior $\dist(\mstate)$ is equivalent to $\sum\nolimits_{\signal \in \signalspace} \joint(\signal,\mstate)$.

When we present a person (hereafter agent) with a decision problem, we give them a choice of response (\textbf{action}) $\action$ from some set of possible responses (\textbf{action space}) $\actionspace$.
The action space can be identical to the state space (as in AI-assisted diagnosis (e.g.,~\cite{alur2024distinguishing}) or can be defined on a different space.
In decision tasks corresponding to prediction tasks, 
(e.g., ``What's the probability the patient will be readmitted to the hospital?''), 
the action space is a probabilistic belief over the state space, i.e., $\actionspace = \distover{\statespace}$.  
The agent faced with a decision task must consider the probability of different states induced by the signal, and choose the best action under the \textbf{scoring rule}, a function that assigns a score representing the quality of the action for the realized state: $S: \Theta \times \actionspace \rightarrow \mathbb{R}$.


Under this definition, the formalization of the joint distribution $\joint$ enables a definition of normative behavior. Summarizing information about $\mstate$, $\joint$ (as well as $\dist$) and communicating it to the participant provides a basis for the experimenter to establish expectations under optimal use of the information. 

\subsection{Normative Behavior}
\label{sec:normative}
Given a decision problem defined as above, we calculate the normative (``optimal'') decision by assuming that the agent has coherent preferences under a scoring rule (dictated by a small set of axioms~\cite{von2007theory}) and uses them to decide between actions under uncertainty about the outcome. 
Consequently, we can interpret an experiment participant's performance as an attempt at achieving this standard, and, if sub-optimal, identify sources of error (loss) in performance.

\subsubsection{Rational Belief Formation}
To define the normative action for a decision problem, we must first characterize how we would ideally expect the agent to form the posterior beliefs $\posterior$ about the state given the signal $\signal$.
A theory of optimal belief construction starts by assuming that the ideal (i.e., rational) agent perfectly perceives $\signal$. 
Once perceived, how does $\signal$ inform the agent's beliefs about $\mstate$? 
Generally, we assume that the agent has some baseline set of beliefs about $\joint$ that they could use to choose an action in the absence of $\signal$. 
Encountering $\signal$ leads the agent to Bayesian update their beliefs. 

Specifically, to calculate the optimal decision for a decision task we first define what the agent perceives the probability distribution over the state to be prior to obtaining information from $\signal$:  $Pr(\mstate)$ or $\dist(\mstate)$ as we describe above.
Whenever the signaling policy does not reveal $\joint(\mstate|\signal)$ directly via the signal, but does inform on $\mstate$, we assume that after seeing the signal, the agent updates their prior belief to a posterior belief over signals and states, using Bayes rule based on what they know about the data-generating model $\joint$. When the signaling policy does reveal $\joint(\mstate|\signal)$ directly via the signal (i.e., $\signalspace \subset \Delta(\Theta)$ and $\signal = \posterior$), this update is trivial.
The maximally informative posterior beliefs that any agent can arrive at given the decision problem is: 

\vspace{-3mm}
\begin{align}
\posterior(\mstate) = \joint(\mstate|\signal) = \tfrac{\joint(\signal,\mstate)}{\sum\nolimits_{\mstate' \in \statespace} \joint(\signal, \mstate')}
\label{eq:posterior}
\end{align}

It is sometimes unreasonable to expect that $\joint$ can be communicated to any agent, rational or not, without incurring some error. For example, in many studies that involve machine learning predictions based on many features and possibly many states (e.g., multi-label classification), communicating the joint distribution over signals and states without loss of information is impractical.  
Under such conditions, we expect the rational agent to arrive at posterior beliefs defined by:

\vspace{-3mm}
\begin{align}
\posterior(\mstate) = \jointapprox(\mstate|\signal) = \tfrac{\jointapprox(\signal,\mstate)}{\sum\nolimits_{\mstate' \in \statespace} \jointapprox(\signal, \mstate')}
\label{eq:posterior_approx}
\end{align}

where $\jointapprox$ represents the approximation of $\joint$ that a rational agent arrives at given the information about $\joint$ employed by the study, such as a sample of labeled instances presented to approximately convey $\joint$.

\subsubsection{Defining the Normative Action}

Given the posterior beliefs as defined in Equation~\ref{eq:posterior} or \ref{eq:posterior_approx}, we use Equation \ref{eq:optimal_decision} to identify the action that a fully rational agent 
would choose in order to maximize their expected utility under $\score$. This is the normative decision.

We assume that the agent's goal is to maximize their expected score under the scoring rule $\score$: 
\vspace{-1mm}
\begin{align}
\score(a,\posterior) &= \expect[\mstate \sim \posterior]{\score(\action,\mstate)} 
\end{align}

\noindent where $\posterior$ describes the agent's belief distribution, the probability distribution over the states that the agent believes the states of the world are drawn from, and $\score(\action, \mstate)$ represents the score assigned by $\score$. 


The optimal action is then the action that maximizes the agent's expected score:
\begin{align}
a_{opt} = \argmax_{\action\in\actionspace} \expect[\mstate\sim\posterior]{\score(\action,\mstate)}
\label{eq:optimal_decision}
\end{align}

\subsubsection{Eliciting and Scoring Beliefs}
A scoring rule induces a ``meaning'' on actions by demarcating a range of beliefs for which an action is optimal. Consequently, when different rules are applied to incentivize participants versus to evaluate their responses, the \textit{meaning} of a given action differs. 
One rule will by definition ``misinterpret'' the actions of the other by implying a different normative standard.
This misinterpretation implies that the scoring rule should be consistent across the two purposes of incentivizing participants and evaluating their responses.

An exception to this requirement occurs when beliefs are elicited instead of decisions, and the rule used to incentivize belief reports has certain properties. Among scoring rules for beliefs, we can distinguish those that are \textit{proper scoring rules}: 
scoring rules with $\actionspace = \distover{\statespace}$ for which the optimal action is to predict
the true distribution, i.e., $\dist \in \argmax_{\action \in \actionspace} \mathbb{E}_{\theta \sim \dist}\score(\action,\theta)$.
Under a proper scoring rule, it is not possible to score higher by reporting some alternative beliefs.
These beliefs can then be plugged into a different proper scoring rule without the risk of misinterpretation.

For any non-proper scoring rule
$\score : \actionspace \times \statespace \to \reals$ there is an
equivalent \textit{proper scoring rule} $\proper : \distover{\statespace}
\times \statespace \to \reals$ defined by playing the optimal action
under the reported belief. 
Formally,
\begin{align}
  \label{eq:proper}
  \proper(\dist,\mstate) &= \score(\argmax\nolimits_{\action \in \actionspace} \score(\action,\dist),\mstate).
\end{align}

For example, applying the construction in Equation \ref{eq:proper} to a binary decision gives a threshold rule where for each state, all reported probabilities below a threshold receive one score and those above receive another score.

\begin{figure*}
    \centering
    \includegraphics[width=\textwidth]{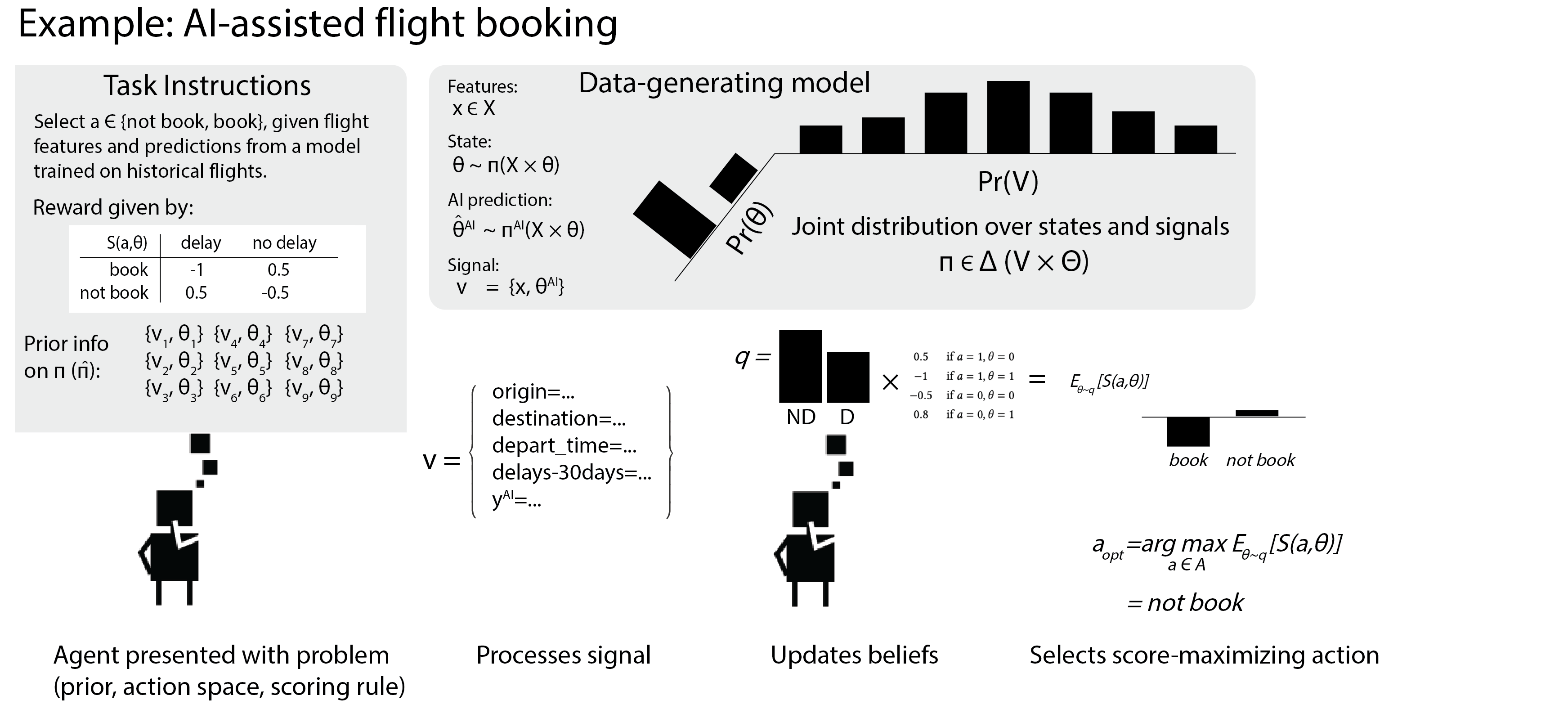}
    \caption{Diagram depicting normative decision for example AI-assisted flight booking scenario. From left to right: The agent is informed of the decision problem, including the action, scoring rule, and prior information about the data-generating model. They next view a signal generated by the data-generating model, which is correlated with the state. The agent updates their beliefs about the state, then chooses the score-maximizing action (in this case, to not book the flight).}
    \label{fig:flight_book}
    \vspace{-5mm}
\end{figure*}

\subsection{Scope of the definitions}
\label{sec:scope}
The above definitions are applicable to any studies that aim to produce \textit{normative} conclusions about human behavior. In such studies, behavioral data (produced by humans or simulation) are collected with the goal of learning about how behavior is induced by the provision of information. Such studies are frequently used to describe the quality of human performance in some situation (e.g., how well people make decisions from displays in strategic settings~\cite{zhang2023designing}), to rank assistive elements by human performance (e.g., different visualizations~\cite{correll2018value}, AI explanation strategies~\cite{liu2021understanding}, or prompts~\cite{buccinca2021trust,hullman2017imagining}), or to test a hypothesis about how humans make decisions~\cite{kale2020visual}.  

It is important to recognize that even if the researchers do not think about the elements of their experiment in terms of the components identified above (i.e., statespace, data-generating model, signaling policy, action space, scoring rule), the design of their study still induces a decision problem whenever two simple conditions hold~\cite{wu2023rational}. First, there must be some uncertainty about the state (i.e., there is more than one data scenario used to generate stimuli)\footnote{A single trial study where all participants view the exact same signal would not qualify.}, and second, it must be possible to score decision quality consistently across participants. This requirement eliminates, for example, studies of emotional responses.

\subsection{Example}
\label{sec:example}

To make these concepts concrete, imagine a study of AI-advised decisions in which people make decisions about whether to book a flight upon seeing features of the flight $\data \in \dataspace$. The state is whether the flight will be delayed beyond some nominal number of minutes, which can take the value 0 for no or 1 for yes ($\mstate \in \{0, 1\}$).
The action space presented to the agent is whether or not to book the flight ($\action \in \{0, 1\}$). 

The data-generating model is the joint distribution over the state (delay or not) and signals comprised of flight feature values and AI prediction information. Assume that participants are presented with a set of feature values describing a flight (origin, destination, time, carrier, prior delays in last 30 days, etc.) from which they must decide on an action. 
Imagine that the signal also includes the prediction from a model that participants are told was trained on a historical dataset of labeled examples.

For a participant to in theory be able to form posterior beliefs about the probability that a flight is delayed after viewing such signals, they must have prior information about the data-generating model. The joint distribution $\joint \in \distover{\signalspace \times \statespace}$ assigns a probability, for each possible value of $\mstate$ (\{0,1\}), to each possible signal $v = \{\data,\hat{\mstate}\}$ (a combination of feature values and model prediction) used as a trial in the experiment.
Rather than attempting to communicate the exact prior to participants, imagine that before beginning the study trials, participants are provided with a finite set of labeled instances $D = \{(\data_1,\hat{\mstate}_1,\mstate_1),...,(\data_m,\hat{\mstate}_m,\mstate_m)\}$, which defines a joint distribution $\jointapprox$ over $V \times \Theta$. Note that if we did not show the agent any feature values at all (nor model predictions), we would expect them to make their choice based on the prior probability that a randomly drawn flight in the study will be delayed, $\jointapprox(\mstate)$.

Imagine the researcher uses a binary scoring rule to map the ground truth realization of the state ($\mstate \in {0,1}$ for no delay, delay) and the participant's selected action from the action space ($\action \in \{0, 1\}$), say, booking the flight ($\action = 1$), to an outcome. 
For example, the agent might be scored by the following rule, which imposes the greatest penalty on false negatives (i.e., booking the flight when it will be delayed):

\vspace{-2mm}
\begin{equation}
    \score(\action, \mstate)=\left\{\begin{array}{ccc}
    0.5 & \text{if }a = 1, \mstate= 0
      & \textit{book, no delay } \\
     -1    & \text{if }\action=1, \mstate=1
     & \textit{book, delay}\\
    -0.5      & \text{if }\action=0, \mstate= 0
    & \textit{not book, no delay}\\
    0.8    &  \text{if } \action=0, \mstate=1
    & \textit{not book, delay}
    \end{array}\right.
    \label{eq:recid_score}
\end{equation}

\noindent where the score could be in units of USD, for example. 

The normative response to this decision problem is defined by the action that a rational agent selects after being presented with the problem (including the action space, scoring rule, and approximation of the data-generating model $\jointapprox$), viewing the signal, updating their beliefs, and selecting the score-maximizing action. Figure~\ref{fig:flight_book} illustrates this example.

\subsection{Defining Sources of Loss}
\label{sec:losses}
Consider the process implied by definitions in Sections \ref{sec:dec_prob}-\ref{sec:normative}, and how each step might result in lost performance relative to a rational agent: 

\vspace{-1mm}

\begin{enumerate}
\item Agent obtains prior $\rightarrow$ prior loss
\item Agent processes signal $\rightarrow$ receiver loss 
\item Agent updates beliefs $\rightarrow$ updating loss
\item Agent takes optimal action $\rightarrow$ optimization loss
\label{eq:process}
\end{enumerate}

For these four sources of loss to explain observed performance losses from study participants, we assume the participants understand the decision problem described to them. However, it always possible that they do not, implying a fifth possible source of loss. 

\vspace{2mm}

\vspace{2mm}
\noindent \textbf{Prior loss (1).}
The agent must obtain prior beliefs that match the prior beliefs the researchers assume in analyzing the problem. 
Prior loss is the loss in performance due to the difference between the agent's prior beliefs and those used by the researchers to calculate the normative standard.


\vspace{2mm}
\noindent \textbf{Receiver loss (2).} The agent processes the signal for the information it provides about the value of the uncertain state.
Receiver loss is the loss due to the agent not properly extracting the information from the signal, for example, because the human visual system constrains what information is actually perceived, because the process of rendering the signal induces noise, or because participants do not understand how to read the signal. 


\vspace{2mm}
\noindent \textbf{Updating loss (3).}
Next the agent updates their prior beliefs with the information obtained.
Updating loss is the loss due to the agent not updating their prior beliefs according to Bayes rule with the information they obtained from the signal, for example due to systematic deviations in human belief-updating relative to rational belief updating as prescribed by Bayes rule (e.g., \cite{camerer1995individual}). 

\vspace{2mm}
\noindent \textbf{Optimization loss (4).}
The agent decides by choosing a good action under their posterior beliefs.
Optimization loss is the loss in performance due to not identifying the optimal action under the scoring rule. 
Optimization loss may include what perceptual psychologists call ``lapse'' responses (e.g., from fatigue or inattention), a catch-all term used to denote responses that do not necessarily reflect the underlying function capturing the person's beliefs. 
Or it may include ``elicitation errors,'' where what participants report on the response scale does not reflect what they had in mind as a result of bias or imprecision induced by the elicitation method.

\vspace{2mm}
\noindent \textbf{Misunderstanding the decision problem.}
A potential contributor to all four sources of loss above is the agent misunderstanding the decision task, including the scoring rule, the statespace, and/or the action space. 
Even when all necessary components of a decision problem 
are carefully communicated, the participant might not read the instructions carefully, or might not understand them.
Consequently, any tests that the researcher runs to validate hypotheses about biases stemming from particular sources of performance loss should be thought of as tests of the joint hypothesis comprised of the target hypothesis \textit{and} the hypothesis that participants understood the task.
\section{Recommendations}
Our primary argument in this work is that to evaluate human decision-making, an experiment must \textbf{provide the participant with enough information to in principle identify the normative decision}, which is used to judge their behavior.
In other words, does the experiment give the participant enough information to align their understanding of the decision problem with its normative interpretation? 

The primary motivations for using the framework are \textit{epistemological} in nature: they concern what knowledge can be gained from the results of a study. 
A lack of consensus around how to define a decision problem and the minimum required components for normative analysis leads to research landscape in which researchers can easily be tempted to overinterpret the results of their studies. Readers may not recognize the overinterpretations, leading to a research literature based on invalid claims about human biases.

In contrast, when the experiment provides the participant sufficient information, we can characterize forms of bias as loss in decision performance according to the four possible sources described in Section~\ref{sec:losses}. When an experiment does not provide sufficient information for the participant to solve the task in principle, there is the potential for multiplicity concerning the data-generating model, as we cannot assume that the participant understood the data-generating model as the researcher intended. 
Consequently, researchers wishing to design a normative decision experiment should communicate all necessary components of the decision problem both to study participants and in publications. Specifically, we recommend that researchers:

\vspace{2mm}
\noindent
\textbf{Clearly convey action and state spaces and a scoring rule.}
The finite action space $\actionspace$ and statespace $\statespace$ should be clearly communicated to participants, and the action space reflected in the instruments used to collect participant responses. 
Because the optimal action can only be determined relative to a scoring rule $\score$ : $\actionspace \times \statespace \to \reals$ that maps the action and state to a quality or payoff, participants should be given a scoring rule to reduce heterogeneity in their choice of action. This is necessary for interpretation even if the experimenter feels comfortable assuming that the participants will try their best and does not use the scores in assigning monetary rewards.

\vspace{2mm}
\noindent
\textbf{Provide prior and/or sufficient information about the data-generating model to calculate posterior.}
As discussed above, 
sufficient information about the data-generating model must be provided to arrive at the Bayesian posterior distribution. 
In designing experiments, researchers should recognize that
there is often more than one way for a study to provide sufficient information for a rational decision-maker to identify the normative decision. However, the fact that experiment designs can vary in the amount of information they provide participants about $\joint$ or how they provide this information does \textit{not} mean that anything goes when it comes to what information is provided. Regardless of the specifics, sufficient information for a rational agent to identify the best response must be provided to interpret the results.

For many experiments, it is critical to convey $\joint$ or an approximation of $\joint$ as the prior in order for participants to be able to update these beliefs upon viewing signals. However, for some experiments, it is instead possible for the signal to directly inform of the posterior probability of the state. For example, in a visualization study that asks users to decide whether or not they want to purchase a piece of special equipment expected to improve their score in a game (e.g.,~\cite{hofman2020visualizing,kale2020visual}), a visualization of the player's expected score distribution with and without the equipment (from which the realized outcome is drawn depending on their choice) conveys the posterior probability of the state directly.

Whenever the signal on its own does not induce a posterior, the prior should be endowed, and sufficient information about $\joint$ should be provided to enable estimating $Pr(\signal|\mstate)$ and $Pr(\signal)$. 
Exemplar-based training procedures such as have been used to teach humans when to defer to an AI model can be employed for this purpose~\cite{mozannar2022teaching}.
More generally, using presentation formats that are more likely to be automatically processed (e.g., conveying distribution information via frequency animations~\cite{goldstein2014lay,hullman2015hypothetical}) or validating participants' priors through elicitation (e.g., ~\cite{kim2019bayesian,kim2020bayesian}) can help researchers understand and minimize prior loss.


\vspace{2mm}
\noindent
\textbf{Use the same scoring rule for incentivization and evaluation, or elicit beliefs with a proper scoring rule.}
\label{sec:score}
As described above, using different scoring rules for incentivizing participants versus evaluating their responses means that ``good'' behavior is defined differently across these two activities, confounding interpretation. Researchers should use the same rule to avoid this confound.

An exception to this rule applies when beliefs are elicited using a proper scoring rule but behavior is analyzed using a different, downstream rule. There are several benefits to eliciting beliefs using a proper scoring rule, in place of or in addition to eliciting decisions. First, when a proper scoring rule is used, the researcher can assume that an agent who understands the rule properly will report their true beliefs. This supports interpretation by eliminating "deviation by design," referring to how deviation from honest reporting 
is not technically a loss under a non-proper rule, it should be the \textit{expected} behavior under that rule. 
In other words, in a non-proper scoring rule, we cannot interpret as biased the behavior of a study participant who does not report the appropriate beliefs after examining the signal.
Eliciting beliefs using a strictly proper scoring rule enables plugging the elicited beliefs into other scoring rules.\footnote{See, e.g., \citet{wu2023rational}'s analysis of \cite{kale2020visual}.}

Second, belief elicitation through proper scoring rules can provide benefits in the form of more information captured in the elicited responses. In general, a coarser action space (i.e., with fewer choices available to participants) will produce less informative responses for the purpose of resolving observed performance relative to a normative standard.
For example, if the researcher designed the flight booking study to also elicit the agent's beliefs, in the form of their prediction of the probability of delay, each trial would provide more information, and they 
could employ a proper scoring rule such as squared loss (i.e., Brier or quadratic rule) or log score to score the accuracy of the probabilistic beliefs.

When eliciting beliefs, one should also acknowledge that different proper scoring rules imply different expectations about the set of possible decisions~\cite{levinstein2017pragmatist}. See \citet{li2022optimization} for a discussion of scoring rules that have a high information value for the rational agent.

\vspace{2mm}
\noindent
\textbf{Compare to best attainable performance.} We observed that even when researchers invest effort to design scoring rules to incentivize participants, they do not often compare the observed behavioral performance to the best attainable performance under the rule. Researchers should default to analyzing results in the space of the scoring rule used for incentives, and provide a rationale whenever they deviate. %

\vspace{2mm}
\noindent
\textbf{Account for learning when providing feedback.}
Some experiment designs support learning about $\joint$ by providing participants feedback informing them of the realized state after each decision trial.
When feedback is provided, we would expect the Bayesian rational decision-maker to update their beliefs according to the information they gain about the joint distribution between signals and states. 
Imagine that participants in the flight booking study see feedback (i.e., are told the realized state, whether or not the flight was delayed) after each decision they make. 
Assuming the signal is not revealing of the posterior probability of the state (delayed or not), we would naturally expect the participants to learn about $\joint$ from this feedback as they complete trials, affecting their posterior beliefs (Equation  \ref{eq:posterior_approx}). 
A challenge in using this approach without providing other information about $\joint$ (i.e., through signals and the prior) is that how much a participant can learn from any single trial is quite limited. Hence, such designs tend to require many trials if the goal is to study behavior under relatively stable beliefs about the data-generating process. 

\vspace{2mm}
\noindent
\textbf{Elicit measures to estimate specific sources of loss.} Assuming the same decision problem has been conveyed to participants as the researchers analyze, it becomes possible to conceive of a set of losses to aid interpretation of results. How to estimate sources of loss in an unbiased way is a rich area for future work. In many cases, we expect separating losses to entail adding carefully constructed questions to the experimental procedure, and/or constructing benchmarks based in rational agent performance to separate sources of loss in observed results. For example, \citet{wu2023rational} use the notion of a calibrated behavioral agent along with a rational agent benchmark and baseline (expected performance of a rational agent under only the prior) to separate optimization and receiver loss.   
An important area for future theoretical work is to establish under what conditions the losses we conceptualize can be estimated.


\vspace{2mm}
\noindent
\textbf{Conduct design analysis.}
Researchers doing normative decision research can benefit from conducting analyses prior to running the study that simulate the idealized actions of an agent to vet the experiment design (see, e.g., \citet{wu2023rational}'s pre-study analyses for visualization studies and \citet{guo2024decision}'s analysis of the potential for complementarity in AI-assisted decisions).
Simulation-based design analysis forces the researcher to explicitly define the parameters of their decision problem, which can elucidate what information needs to be communicated to a participant in order to enable an optimal response.
It also enables the researcher to test variations on their data-generating model and scoring rule, which can help the researcher realize under what conditions (1) the experiment can have acceptable statistical power and (2)  participants can be paid fairly.
Design analysis helps the researcher choose conditions (e.g., AI explanations) that provide informative signals for the decision task, narrowing down what kinds of research questions can be answered using a particular decision problem.
For example, at a minimum, to be informative of how effectively people use signals, a study design must have sufficient statistical power to separate the rational agent performance with and without the signal.

\section{Examples: Using the Framework to Redesign Confounded Studies}
\label{sec:examples}
We use several examples inspired by the human-centered AI and visualization literature to demonstrate why drawing conclusions about biases in human behavior is misleading in scenarios where sufficient information for a rational agent to optimize is not provided. We discuss how the example study designs could be improved to align with the definitions in Section~\ref{sec:def}. 

\subsection{AI-Assisted Flight Booking}
We first elaborate the flight booking study mentioned earlier. Imagine the study is designed to investigate whether people over-rely on AI assistance when deciding whether to book a flight or not.
The researchers use a standard train-validate-split approach on a large labeled historical data set to train a model that predicts whether or not a flight will be delayed.  
Each flight in the dataset is represented by a set of feature values, $\dataRV=\data$, which comprise part of the signal $\signal$ shown on each trial.
Study participants are randomly assigned to either a control condition where they see the feature values along with only the model's prediction and what is described as ``the model's confidence in its prediction, ranging between 50\% and 100\%,''  or an AI explanation condition where they are also given additional information about how features in the data contribute to the model's predictions, e.g., generated by LIME~\cite{ribeiro2016should}.
The model confidence score that is provided is an approximately calibrated probability\footnote{A confidence function $f_{B}: X \rightarrow [0,1]$ is $\alpha$-calibrated with respect to a sample $S \subseteq X$ if there exists some $S' \subseteq S$, with $|S'| \geq (1-\alpha)|S|$, such that for any $b \in [0,1]$: $|P(Y=1| f_{B}(X)=b, X \in S') - b| \le \alpha$~\cite{corvelo2024human}. In other words, the probabilities are approximately calibrated with respect to a sample $S$ when for some subset at least $1-\alpha$ times the size of $S$, the difference between the stated and real probability is less than or equal to $\alpha$.} 
of a correct prediction conditional on the feature values, $Pr(correct|\dataRV=\data)$. 


Participants are told the AI model was trained on prior flights whose delay status is known.
They are told that the model has an accuracy of $80\%$ (representing the combined false positives and false negatives) on a sample of instances drawn from the test set which comprise the trials in the study.

Each participant makes booking decisions for 20 unique flights. On each trial they are asked to choose whether or not to book the flight ($\action \in \{0,1\}$).
Rather than the scoring rule given in Equation \ref{eq:recid_score}, imagine a simpler version that assigns different types of errors equivalent costs. Participants are given a base reward of $\$10 USD$, but they are told they will lose $\$0.5 USD$ from this initial endowment for every trial where they either book a flight that will be delayed or do no book a flight that will not be delayed.
Participants are \textit{not} told how often to expect delays; i.e., the prior $\dist$ is not endowed, nor is $\joint$.
Participants are \textit{not} given interim feedback on their decisions.

After collecting data from 500 participants, the researchers model the percentage of trials where the participants incorrectly follow the AI's advice.
They find a statistically significant difference between the AI explanation condition and the control condition, with the rate of incorrectly following the AI's advice being higher in the former.
The researchers interpret this as evidence that AI explanations promote over-reliance.

\vspace{2mm}
\noindent \textbf{Interpretation of Results.}
To assess what can be said about performance loss, we first consider whether the problem is well-defined: could a rational agent use the provided information to select the utility-maximizing decision?

The action space $A$ ($\action=1$ for book, $\action=0$ for not book) is conveyed to participants, as well as the state space $\statespace$ ($\mstate=1$ for the flight being delayed, $\mstate=0$ for not). 
Through the information they are given about the task payouts, they also have knowledge of the scoring rule that assigns a score (reward amount) given the realized state and their decision.

How would a rational participant form beliefs about the posterior probability of the state? 
Participants are not provided with information about the prior probability of the state.
This would be acceptable if the signal provided them with posterior beliefs about the uncertain state. However, the signal does not provide the required posterior probability: recall that participants receive the estimated probability that the model is correct conditional on the features, $Pr(correct|\dataRV=\data)$, but not conditional on the model's prediction. They therefore cannot obtain sufficient information from the data-generating model $\joint$ to guide their choice of action. 


\vspace{2mm}
\noindent \textbf{Improving the experimental design.}
One possibility is to make the signal provide the posterior $\posterior(\mstate)$.
This could be achieved if the confidence value included in the signal were to give the probability that the model is correct conditional on both the features and the prediction, e.g., $Pr(correct|\dataRV=\data,y_{AI}=1)$ when the model predicts 1. 
In this setting, provision of the prior would not be necessary (and may confuse participants).

However, the careful reader may also have noticed another barrier to defining bias for the problem: the model's conditional confidence is described as simply ``the model's confidence in its prediction'', making it ambiguous from the participant's perspective whether it is conditional on the features and model prediction or only one of the two.
The normative response is not well-defined if the probability is not conditional on both, unless the researchers also communicate $\joint$ or an approximation $\jointapprox$ to participants before they begin the decision trials.  
Hence the model probability description would need to be amended to specify that the confidence is conditional on the predicted label to eliminate ambiguity.

It is worth noting that decision problems do not necessarily have to depend on small, discrete action spaces. 
Imagine another amendment to the flight booking study where instead of asking participants for a booking decision, the researchers ask them to report the probability that the flight is delayed $Pr(\mstate|\signal)$. 
Assuming the above changes are made to the design to equip participants with information to form posterior beliefs, participants could in principle report the optimal beliefs. 
However, if the scoring rule for a decision problem where the action space is over beliefs is not proper, we should not expect the agent to respond with their true beliefs.
Optimization loss might be low, but this is not useful to the experimenters' research questions if agents respond with a distortion of their true beliefs. 

\subsection{Election Forecast Displays}
Imagine a study targeting the effect of different ways of visualizing election forecasts. The researchers develop a forecasting model to predict vote share (a percentage) by weighting and aggregating poll results from news sources for ten unique upcoming two-party state-level elections. Participants are randomly assigned to view one of three forecast displays showing the predicted vote share with uncertainty of each party's candidate as a table, an animated graphic, or a static graphic. Assume for simplicity that all three induce informationally equivalent signals to a rational agent\footnote{If any of the displays was not informationally equivalent to the others, the optimal decision may vary by display. See Wu et al.~\cite{wu2023rational} for analysis of such examples using a rational agent framework.}.

Upon beginning the experiment, participants are randomly assigned to a display treatment and also assigned a political party preference.
They are told they will view election forecasts from a model based on poll results and should imagine they live in the district where the election takes place. 
Each participant in the study does ten trials corresponding to the ten elections. 
On each trial, they are asked if they would like to vote for the candidate in the election. Voting carries no cost. 

Participants are told they will win a given base reward amount plus an additional $\$0.25 USD$ ``for each trial where their candidate wins a simulated version of the election that uses the forecast model to predict how people will vote.''
The researchers generate
outcomes by randomly drawing a bivariate vote share outcome from the predicted distribution under the model and generating the uncertain state $\theta$ (taking the value 0 or 1 for lose or win) by assuming the higher vote share candidate wins.
Participants are \textit{not} given feedback after each trial.

After collecting results for several hundred participants, the researchers fit a linear model that estimates the sensitivity of the vote decision to a variable denoting the closeness of the race, operationalized as the difference between the predicted vote share of the first candidate and 0.5. The researchers postulate that a better display induces more sensitivity.
Using statistical significance testing, they reject the null hypothesis of no difference between the estimated slopes for the interaction between display type and race closeness across display types. Using posthoc comparisons, they declare that the animated graphic leads to better voting decisions because it induces the most sensitivity. 

\vspace{2mm}
\noindent \textbf{Interpretation of Results.} 
To understand why the researchers' conclusions might be misleading as a prediction of real world use of election displays, consider the decision problem from the perspective of the participants.
The action space $\actionspace$ ($\action=1$ for vote, $\action=0$ for not vote) is conveyed to participants, and the task description also gives participants the statespace $\statespace$ ($\mstate=1$ for their candidate winning, $\mstate=0$ for not). 
Participants can infer the probability of their candidate winning 
from the signal showing the model predicted vote share. Because the signal provides the posterior probability of the state, no prior information needs to be communicated.

What do participants know about the realized outcome of the election that the researchers describe simulating? 
The researchers described the use of the forecast model to simulate voting behavior, and in implemention, do \textit{not} account for the participant's vote in generating the outcome. Consequently, whether a participant votes is irrelevant to the payoff-relevant state under the scoring rule.
However, the description provided to participants is ambiguous about whether the participant's vote counts. Some participants might assume, based on their knowledge of real-world elections, that it does. Hence performing well in the experiment requires being able to guess what the researchers have in mind, rather than behaving rationally in the context of the experiment. 

Because, in the context of the study, voting carries no value but also no cost, we cannot conceive of the four sources of loss. There is no well-defined decision problem, and consequently, no formal justification for concluding that behavior is more biased or otherwise worse under one display condition.

\vspace{2mm}
\noindent \textbf{Improving the experimental design.}
We could create a valid decision problem by changing the scoring rule so that the reward amount depends on whether one votes. For example, we could assign a cost of $\$0.25$ to voting:


\vspace{-2mm}
\begin{equation}
   \score(\action, \mstate)=
    =\\ \left\{\begin{array}{ccc}
    0 & \text{if }a = 0, \mstate= 0
      & \textit{do not vote, lose election } \\
     0.5    & \text{if }\action=0, \mstate=1
     & \textit{do not vote, win election}\\
    -0.25      & \text{if }\action=1, \mstate= 0
    & \textit{vote,  lose election}\\
    0.25    &  \text{if } \action=1, \mstate=1
    & \textit{vote, win election}
    \end{array}\right.
    \label{eq:vote_score}
\end{equation}

However, the decision problem does not admit a well-defined normative response until we address the ambiguity in the description of the data-generating model. 
Only if participants are given a transparent description of how the realized outcome is generated from the model can we conceive of prior loss as well as updating loss, both of which should be zero given the signal design. We can conceive of receiver loss as the loss from participants not being able to extract the information from the signal, and optimization loss as the loss from participants not being able to choose the optimal action under their beliefs. 

Of course, if we amend the description to inform participants of how the election outcome is actually generated, we should expect participants who understand the problem to never vote. Rather than informed political engagement, the proportion of participants who vote (e.g., in each treatment condition) could be described as reflecting the level of confusion the participants had about the task.
Or, if the researchers did not change the description of how the realized state is generated, but designed the study to give participants feedback on the realized state after each decision, then the participants might in theory learn that their vote does not matter. This would work against the researchers' goals of presenting a situation where people could feel compelled to vote.

This example demonstrates that providing participants sufficient information to determine the utility-maximizing action is necessary but not sufficient for a meaningful decision problem in light of the research questions of interest.  

\section{Applicability of the Normative Framework to HCAI Studies}
To illustrate the applicability of our definition, we present a meta-study in which we analyzed a sample of recent evaluative studies of AI-assisted decisions to estimate how often researchers draw conclusions about flaws in human decisions without a well-defined decision problem. Note that HCAI studies are just one example of the kinds of decision studies to which our work applies (Section~\ref{sec:scope}).

\vspace{2mm}
\noindent \textbf{Sample.}
We sampled 46 of the studies surveyed by \citet{lai2023towards} in their overview of empirical human-subjects studies on human-AI decision-making. According to their inclusion criteria, the sample contains only evaluative human-subject studies targeting a decision task in the context of classification or regression published in ACM or ACL conference venues between 2018 and 2021. 

\vspace{2mm}
\noindent \textbf{Codes.} In coding the studies, we focused on three aspects of the study design and interpretation. First, we identified whether our framework is \textit{applicable}: Does the task given to participants admit a ground truth state \textit{and} induce uncertainty about the state, at least across participants? Next, we identified whether the authors draw \textit{conclusions about flaws or performance loss in human decisions} from their results, i.e., is the study evaluating human decision-making in some way? We identified whether the study targeted \textit{a well-defined problem}; i.e., whether participants in the study were provided with sufficient information for a rational agent to best respond to the task.
Finally, we noted whether the researchers \textit{reported conclusions using a different scoring rule} than the one provided to participants, and whether participants were incentivized using a \textit{flat scoring rule} (i.e., a rule that does not distinguish between responses based on their distance to the ground truth). Note that studies that pay participants a flat reward require that researchers report conclusions using a different scoring rule, introducing confounding.

In some cases, coding was challenging because the paper did not provide much detail on what exactly was communicated to participants, nor provide the study interface as a supplement. This demonstrates a lack of sensitivity on the researchers' part when it comes to providing requisite information to interpret their results. We noted such cases, and discuss how specific omissions are problematic. 


Two authors participated in the coding. After formulating the coding scheme, the authors divided the sampled studies roughly evenly, each coding independently, then discussing all challenging examples to ensure consensus.
The list of coded studies, our coding guide, and full coding results are available as an Appendix.

\vspace{2mm}
\noindent \textbf{Results.} 
Of the 46 studies examined, 39 (85\%)
studied tasks to which the framework was applicable. One study posed questions and drew conclusions about the diagnostic utility of AI recommendations, but elicited only perceived utility~\cite{cai2019human}. We coded this example as Applicable. The remaining 7 (15\%) included studies relegated to perceptions (e.g., of fairness, accuracy) and of subjective appraisals of music or movie recommendations or emotion.

All 39 of the applicable studies drew conclusions about flaws in human decisions, such as by pointing to over- or underreliance on the AI or speculating about why participants failed to get the correct answer. 
However, when we evaluated whether participants were given sufficient information to identify the best decision response across all conditions, this was true in only 7 of the 39 (18\%) cases. 
In another 3 of 39 (7.7\%), participants were given sufficient information to identify the best decision response in some subset of conditions.
In the latter cases, informational differences between conditions that the researchers compared could lead to different a priori expectations of decision quality, yet these differences were not accounted for in comparing participants' behavior between conditions\footnote{The calculation of separate performance benchmarks would address such ambiguity~\cite{wu2023rational}.}. 
While the framework was clearly applicable, there was not enough information to determine if participants were presented with a well defined problem in 4 (10\%) of the remaining studies.
In other words, at best, over 70\% of the surveyed studies did not give participants an opportunity to do well by the evaluation approach the researchers applied.

Only 7 (18\%) of the 39 studies to which the framework is applicable used the same scoring rule to analyze the responses as was used to incentivize participants, illustrating a lack of awareness of the implications of using different rules for interpreting the meaning of responses. 25 (64\%) did not, one did in at least one experiment, and another 6 (15\%) did not report enough detail for us to know.
Finally, 14 (36\%) of the 39 used a flat rule for incentives, one did in some experiments, 16 (36\%) did not, and 7 (18\%) did not report enough detail for us to know.

We observed that even when authors gave participants sufficient information to identify the best response, they rarely compared the observed performance to the best attainable performance.
For example, multiple papers defined a scoring rule to incentivize participants to make accurate decisions, but evaluated the responses using a simpler accuracy scale, implying that 100\% accuracy was obtainable even when it would not be to a perfectly-perceiving rational agent. 
In some cases, participants were also given sufficient information to optimize, but it did not accurately reflect the true data-generating model, such that the results interpretation did not hold. The most common scenario was when the researchers gave participants information about the model's held-out accuracy but the participants experienced instances where the model had a different accuracy. Especially when the study design includes feedback, this leads to a contradiction between the instructions and what participants observe which makes it difficult to draw any evaluative claims about behavior.

Altogether, these results suggest that the framework is widely applicable to swaths of studies currently being conducted in HCI and related fields. Moreover, they suggest that researchers stand to improve the interpretability of their research by adopting it by following recommendations that it implies, as we describe further below.

\begin{figure*}
    \centering
    \includegraphics[width=0.6\textwidth]{chi2025_charts.png}
    \caption{Results from coding 46 studies surveyed by \citet{lai2023towards}. Seven studies did not evaluate human decisions, instead focusing on capturing perceptions or subjective appraisals. Of the remaining 39, 25 did not communicate sufficient information to participants for them to identify the best response. Additionally, 25 did not use the same scoring rule for incentivizing participants as for analyzing their responses.}
    \label{fig:coding_results}
    \vspace{-5mm}
\end{figure*}

\section{Discussion: Goals and Incentives in the Lab versus the World}
By arguing that researchers adopt and sufficiently communicate to participants a widely-applicable definition of a decision problem, our goal is to improve the integrity of claims made from human subjects experiments. 
When participants are not given ample information to ``solve'' at least in principle the decision problem, we cannot bring well-formed expectations to the results, and should expect heterogeneity in the behaviors they exhibit.
Such ``heterogeneity by design'' renders the results uninterpretable for evaluative purposes like identifying forms of bias.


Imagine that the researchers who designed the voting task intentionally did not include the vote in the scoring rule, and were intentionally ambiguous about the role of the vote in the data-generating model. 
Because the utility of voting in a real-world election is typically known only implicitly at an individual level and likely to vary between people, the researchers might think that their results would therefore be more
representative
of real world voting behaviors. 

Such beliefs stem from a convention among many researchers toward being optimistic about how their experimental observations relate to people’s behavior in the ``real world.'' Most controlled decision studies are designed to help us understand how people behave in some realistic setting that the experimental task is meant to proxy for. However, 
in the real world, people have goals and prior beliefs. We might not be able to perceive what utility function each individual person is using, but we can assume that behavior is goal-directed in one way or another. Savage’s axioms~\cite{steele2015decision}
and the derivation of expected utility theory tell us that for behavior to be rationalizable, a person’s choices should be consistent with their beliefs about the state and the payoffs they expect under different outcomes.

Consequently, when people are in an experiment, we should not expect the analogous real world goals and beliefs for that kind of task to apply. 
For example, people might take actions in the real world for intrinsic value---e.g., voting in order to feel like a good citizen, or paying close attention to features of a flight they are deciding whether to book because the decision will impact whether or not they arrive in time for an important event. 
It is difficult to motivate people to take actions based on intrinsic value in an experiment, unless the experiment is designed to elicit and describe those intrinsic values, in which case the normative framework here no longer applies.
Hence, the idea that as experimenters we can skip fully describing the decision problem to our participants while still retaining the ability to identify their behavior as faulty or biased arises from a failure to recognize our fundamental uncertainty about how the experimental context relates to the real world. 

\subsection{Implications for Research Ethics}
The requirement of communicating to study participants sufficient information for a rational agent to optimize may seem unnecessarily strict if one assumes that participants' domain knowledge alone should enable them to evaluate the information about the state captured in signals.
For example, in the flight booking example, a researcher may be tempted to assume that even without being given any information about the probability that the model prediction is correct, a participant will be able to use their prior knowledge about features of delayed flights to infer the probability that the model prediction is correct. 
The problem is that once a researcher makes this assumption, they are no longer evaluating participants on only the information they provided. They are also implicitly evaluating them on how well their prior beliefs about the state and the validity of the model's learned information align with the researcher's choice of data-generating model. There are various implications of this misalignment that we may find problematic upon reflection. For example, under such conditions, participants must bring identical beliefs to the researchers' in order to have a chance of scoring well. 
A related point has been made in research on crowdsourcing, where it has been observed that there are often many plausible ways to interpret a crowdsourced task based on the information workers are provided, but work requesters often use quality control algorithms that recognize and pay only one of these~\cite{alkhatib2019street,kairam2016parting}. 

Fortunately, this does not mean that human decisions can only be evaluated by assuming that participants bring no additional information. Instead, the data-generating model and decision task should account for the presence of human prior information. For example, for AI-assisted decision tasks like the flight booking example (\ref{sec:example}), a human prediction of the delay status of a flight, governed by a distinct generating process from the AI prediction, can be included in the data-generating process. The decision problem for which the normative decision is sought can then be cast as a decision between the human and AI prediction. If independent human predictions are not available, they can be elicited from the subjects before providing the signal. Recent work by Guo et al.~\cite{guo2024decision} demonstrates this approach.



\subsection{Limitations \& Future Work}
\label{sec:limits}
Our argument is that the interpretation of results is restricted when a well-defined decision problem is not conveyed to participants, not that researchers cannot conduct comparisons at all without a well-defined decisions. Researchers can conduct and report on studies that do not conform to this approach, but they cannot conclude that people made bad decisions if they were not given sufficient information to know what a good decision is. For example, imagine a study comparing the decision performance of users with and without the use of recommendations from an AI, where no prior is given to participants, and no scoring rule is given. Imagine that performance in the study is measured in terms of accuracy, and average accuracy is higher among participants who had access to the AI. Given an appropriately designed randomized experiment, the researcher could conclude that there is a causal effect of the AI on decisions. However, they could not claim that people are making worse or biased decisions without the AI, because they are scoring their decisions using information they never provided to participants. 
When we say that one condition did better, we should acknowledge that we did not give them a chance to fully understand and optimize their behavior for the problem that we analyzed.

A common misinterpretation of the use of a rational agent model is that we are implying that human behavior is expected to be perfectly rational, or approximately rational.
This is not the case. While a rational agent approach formally justifies what it means to best use information to inform the selection of an action using statistical decision theory, participants in our experiment may use different strategies. The point is that without a definition of what it means to best respond to the information we have provided, we are hard-pressed to evaluate participant behavior. Even if we do not expect this procedure to characterize what human agents do, it provides a basis for interpreting the results in terms of deviations from a well-understood standard.

Several challenges have been raised against
expected utility theory as put forth by Savage~\cite{savage1972foundations}; we recommend \cite{steele2015decision} for an overview. 
Another critique of a normative approach dubbed ``value collapse''~\cite{nguyen2023value} is that presenting a clear set of values as a target for behavior limits variance in ways that can be harmful. 
Our argument is that we should not mistake the variation that emerges from in an under-communicated decision problem as representative of behavior under a particular set of values or a particular real-world decision task. 
If a researcher wishes to evaluate decisions but contests Bayesian decision theory as a normative standard, they should be prepared to motivate whatever standard they adopt instead with equivalent rigor (e.g., derivation from interpretable assumptions combined with mathematical laws). We note that alternative models have been well-explored, but rarely achieve provable grounding relative to mathematical laws as statistical decision theory does.

Sometimes researchers are explicitly interested in understanding the range of values at play in a naturalistic decision scenario rather than providing a well-defined decision problem. In this case, their research is better characterized as descriptive, and they should refrain from claims that hinge on evaluating participant decisions.
While our framework is intended for normative decision research, the definitions may also be useful if one is doing such descriptive or exploratory decision research, by making clear which aspects of the decision problem are underspecified and therefore presumably the object of study. For example, a researcher might aim to characterize challenges in model-assisted decisions in a particular domain~\cite{zytek2021sibyl}, or conduct an exploratory study to elicit information about the utility function (i.e., scoring rule) that operates in a setting and then communicate it back to new participants as part of the decision problem in an evaluative experiment (e.g.,~\cite{fernandes2018uncertainty}).

 
Some have voiced pessimistic views on the value of monetary incentives for experiment participants (e.g.,~\cite{read2005monetary}). 
As \citet{wu2023rational} show, this view is to be expected under a rational agent approach whenever the value of information to the decision problem (i.e., the difference in the rational agent's expected score given only the prior versus with access to the signals~\cite{wu2023rational}) is small. 
This motivates analyzing whether a decision study is likely to provide agents with a sufficient incentive to motivate them to consult the signals~\cite{wu2023rational}. 

Of course, scoring rules can sometimes be challenging for study participants to interpret. 
Theoretic incentive compatibility does not guarantee understanding on the part of subjects. 
There is a rich literature discussing possible descriptions that can be used with common rules like the quadratic scoring rule (e.g.,~\cite{artinger2010applying}). Interfaces researchers may be in a good position to extend this literature through careful observational studies.
More broadly, however, the insensitivity of human participants to monetary incentives are orthogonal to the crux of our argument, which is philosophical and concerned
with what is knowable more so than the pragmatics of how much behavior changes when a problem is well-defined.


\section{Conclusion}
We present definitions of a decision problem and normative response synthesized from statistical decision theory and information economics for the purpose of improving the interpretability of decision studies in HCI, visualization, human-centered AI, and related fields. We demonstrate the applicability of the framework to recent HCAI studies, and provide specific recommendations to researchers and examples of how confounded study designs can be made interpretable or well-formed within the framework. 
Ultimately, the goal of our work is to give researchers the tools to recognize that when we do experiments, we are restricted to learning about behavior in the artificial worlds we create. As much as we might want to equate our results with some real world setting used as motivation, extrapolating from the world of the controlled experiment to the real world will always be a leap of faith. It is therefore the responsibility of the experimenter to fully understand their experimental world, and, for the purposes of offering the results as a form of potentially generalizable knowledge, to give their participants a chance to do so as well. 


\begin{acks}
We thank Yifan Wu and Ziyang Guo for discussions on these topics, and Dan Weld and Matt Groh for feedback on a draft.
\end{acks}

\bibliographystyle{ACM-Reference-Format}
\bibliography{decisions}

\section{Appendix}

\subsection{List of coded studies}
\label{list_of_studies}

We randomly sampled and coded the following studies from \citet{lai2021towards}:

1. A. Abdul, C. von der Weth, M. Kankanhalli, B. Lim. (2020). COGAM: Measuring and Moderating Cognitive Load in Machine Learning Model Explanations.

2. A. Alqaraawi, M. Schuessler, P. Weiß, E. Costanza, N. Berthouze. (2020). Evaluating Saliency Map Explanations for Convolutional Neural Networks: A User Study.

3. G. Bansal, T. Wu, J. Zhou, R. Fok, B. Nushi, E. Kamar, M. T. Ribeiro, and D. Weld. (2021). Does the whole exceed its parts? the effect of AI explanations on complementary team performance. 

4. G. Bansal, B. Nushi, E. Kamar, W.S. Lasecki, D.S. Weld, E. Horvitz. (2019). Updates in Human-AI Teams: Understanding and Addressing the Performance/Compatibility Tradeoff.

5. G. Bansal, B. Nushi, E. Kamar, W.S. Lasecki, D.S. Weld, E. Horvitz. (2019). Beyond Accuracy: The Role of Mental Models in Human-AI Team Performance.

6. R. Binns, M. Van Kleek, M. Veale, Ulrik Lyngs, J. Zhao, N. Shadbolt. (2018). ‘It’s Reducing a Human Being to a Percentage’; Perceptions of Justice in Algorithmic Decisions.

7. Z. Buçinca, M.B. Malaya, K.Z. Gajos. (2021). To Trust or to Think: Cognitive Forcing Functions Can Reduce Overreliance on AI in AI-assisted Decision-making.

8. Z. Buçinca, P. Lin, K.Z. Gajos, E.L. Glassman. (2020). Proxy Tasks and Subjective Measures Can Be Misleading in Evaluating Explainable AI Systems.

9. A. Bussone, S. Stumpf, D. O'Sullivan. (2015). The Role of Explanations on Trust and Reliance in Clinical Decision Support Systems.

10. C. Cai, E. Reif, N. Hegde, J. Hipp, B. Kim, D. Smilkov, M. Wattenberg, F. Viegas, G.S. Corrado, M.C. Stumpe, M Terry. (2019). Human-Centered Tools for Coping with Imperfect Algorithms during Medical Decision-Making.

11. S. Carton, Q. Mei, P. Resnick. (2020). Feature-Based Explanations Don't Help People Detect Misclassifications of Online Toxicity.

12. A. Chandrasekaran, V. Prabhu, D. Yadav, P. Chattopadhyay, D. Parikh. (2018). Do Explanations make VQA Models more Predictable to a Human?

13. H-F. Cheng, R. Wang, Z. Zhang, F. O'Connell, T. Gray, F.M. Harper, H. Zhu. (2019). Explaining Decision-Making Algorithms through UI: Strategies to Help Non-Expert Stakeholders.

14. M. Chromik, M. Eiband, F. Buchner, A. Krüger, A. Butz. (2021). I Think I Get Your Point, AI! The Illusion of Explanatory Depth in Explainable AI.

15. D. Das and S. Chernova. (2020). Leveraging Rationales to Improve Human Task Performance.

16. B.J. Dietvorst, J.P. Simmons, C. Massey. (2018). Overcoming Algorithm Aversion: People Will Use Imperfect Algorithms If They Can (Even Slightly) Modify Them.

17. J. Dodge, Q.V. Liao, Y. Zhang, R. Bellamy, C. Dugan. (2019). Explaining Models: An Empirical Study of How Explanations Impact Fairness Judgment.

18. J. Dressel, H. Farid. (2018). The accuracy, fairness, and limits of predicting recidivism.

19. K. Gero, Z. Ashktorab, C. Dugan, Q. Pan, J. Johnson, M. Ruiz, S. Miller, D. Millen and W. Geyer. (2020). Mental Models of AI Agents in a Cooperative Game Setting.

20. B. Ghai, Q.V. Liao, Y. Zhang, R. Bellamy, K. Mueller. (2021). Explainable Active Learning (XAL): An Empirical Study of How Local Explanations Impact Annotator Experience.

21. B. Green, Y. Chen. (2019). Disparate Interactions: An Algorithm-in-the-loop Analysis of Fairness in Risk Assessments.

22. S. Guo, F. Du, S. Malik, E. Koh, S. Kim, Z. Liu, D. Kim, H. Zha, and N. Cao. (2019). Visualizing Uncertainty and Alternatives in Event Sequence Predictions.

23. P. Hase, M. Bansal. (2020). Evaluating Explainable AI: Which Algorithmic Explanations Help Users Predict Model Behavior?

24. R. Kocielnik, S. Amershi, P.N. Bennet. (2019). Will you Accept an Imperfect AI? Exploring Designs for Adjusting End-user Expectations of AI Systems.

25. T. Kulesza, S. Stumpf, M. Burnett, S. Yang, I. Kwan, W-K. Wong. (2013). Too Much, Too Little, or Just Right? Ways Explanations Impact End Users’ Mental Models.

26. J. Kunkel, T. Donkers, L. Michael, C-M. Barbu, J. Ziegler. (2019). Let Me Explain: Impact of Personal and Impersonal Explanations on Trust in Recommender Systems.

27. I. Lage, E. Chen, J. He, M. Narayanan, B. Kim, S. Gershman, F. Doshi-Velez. (2019). An Evaluation of the Human-Interpretability of Explanation.

28. A. Levy, M. Agrawal, A. Satyanarayan, D. Sontag. (2021). Assessing the Impact of Automated Suggestions on Decision Makingg: Domain Experts Mediate Model Errors but Take Less Initiative.

29. V. Lai, C. Tan. (2019). On Human Predictions with Explanations and Predictions of Machine Learning Models: A Case Study on Deception Detection.

30. Z. Lin, J. Jung, S. Goel, J. Skeem. (2020). The limits of human predictions of recidivism.

31. J.M. Logg, J. Minson, D.A. Moore. (2019). Algorithm Appreciation: People Prefer Algorithmic To Human Judgment.

32. Lu, Z., M. Yin. (2021). Human reliance on machine learning models when performance feedback is limited: Heurstics and risks.

33. K. Mallari, K. Inkpen, P. Johns, S. Tan, D. Ramesh, E. Kamar. (2020). Do I Look Like a Criminal? Examining how Race Presentation Impacts Human Judgement of Recidivism.

34. S. McGrath, P. Mehta, A. Zytek, I. Lage, H. Lakkaraju. (2020). When Does Uncertainty Matter?: Understanding the Impact of Predictive Uncertainty in ML Assisted Decision Making.

35. D. Nguyen. (2018). Comparing Automatic and Human Evaluation of Local Explanations for Text Classification.

36. J.S. Park, R. Barber, A. Kirlik, K. Karahalios. (2019). A Slow Algorithm Improves Users' Assessments of the Algorithm's Accuracy.

37. F. Poursabzi-Sangdeh, D.G. Goldstein, J.M. Hofman, J. Wortman Vaughan, H. Wallach. (2021). Manipulating and Measuring Model Interpretability.

38. E. Rader, K. Cotter, J. Cho. (2018). Explanations as Mechanisms for Supporting Algorithmic Transparency.

39. A. Smith-Renner, R. Fan, M. Birchfield, T. Wu, J. Boyd-Graber, D. Weld, L. Findlater. (2020). No Explainability without Accountability: An Empirical Study of Explanations and Feedback in Interactive ML.

40. A. Springer, S. Whittaker. (2019). Progressive Disclosure: Designing for Effective Transparency.

41. K. Stowers, N. Kasdaglis, M. Rupp, J. Chen, D. Barber, and M. Barnes. (2017). Insights into Human-Agent Teaming: Intelligent Agent Transparency and Uncertainty.

42. M. Szymanski, M. Millecamp, K. Verbert. (2021). Visual, textual or hybrid: the effect of user expertise on different explanations.

43. N. van Berkel, J. Goncalves, D. Russo, S. Hosio, M.B. Skov. (2021). Effect of Information Presentation on Fairness Perceptions of Machine Learning Predictors.

44. X. Wang, M. Yin. (2021). Are Explanations Helpful? A Comparative Study of the Effects of Explanations in AI-Assisted Decision-Making.

45. M. Yin, J. Wortman Vaughan, H. Wallach. (2019). Understanding the Effect of Accuracy on Trust in Machine Learning Models.

46. K. Yu, S. Berkovsky, R. Taib, J. Zhou, F. Chen. (2019). Do I trust my machine teammate?: an investigation from perception to decision.

\subsection{Coding guide}
\label{coding_guide}
Our coding approach is designed to separate the following questions:
\begin{itemize}
\item Does the study include a task for which a ground truth response can be defined?
\item Does the study give the participant a decision problem that a rational agent could best respond to?
\item Do the researchers interpret the results assuming the same decision problem given to participants, or do they assume a different problem (e.g., by changing the scoring rule when they evaluate responses)?
\end{itemize}

Through iterative collaborative coding, the first two authors developed five codes:
\begin{enumerate}
    \item \textbf{Applicability of framework}: Do the research questions concern a type of task for which a ground truth state can be defined?
    \item \textbf{Evaluative conclusions}: Do the authors draw \textit{conclusions about flaws or performance loss in human decisions} from their results, such as ranking different conditions, or implying that participants could have done better in any condition? 
    \item \textbf{Well-defined problem}: Does the task presented to participants represent \textit{a well-defined decision problem}? In other words, are participants in the study were provided with sufficient information for a rational agent to best respond to the task?
    \item \textbf{Same rule (problem) reported?} Do the authors interpret the results assuming the same scoring rule given to participants, such that the results bear on the same decision problem given to participants? 
    \item \textbf{Flat rule?} Does the rule used to incentivize participants assign the same score to all responses regardless of the ground truth? 
\end{enumerate}

When the framework was not applicable, we did not proceed to code the remaining items. In cases where not enough information was provided in the paper text or the supplemental material to determine the code, we noted this. When papers reported multiple studies or multiple conditions in the same study that differed in their codes, we used reported Sometimes and Partially, respectively.

We coded the third item above--Is the problem given to participants well-defined?--by requiring that the following three criteria were fulfilled:
\begin{itemize}
\item Does the task given to participants admit a ground truth state and induce uncertainty about the state, at least across participants?
\item Are participants provided with ample information about the data-generating model for a rational agent to understand how to best respond (i.e., optimize)? 
\item Are participants provided with a scoring rule that unambiguously assigns a score to each possible response in the action space?  
\end{itemize}

Note that the first criteria is distinct from Applicability of the framework code described above: the former asks about the task given to participants, while the latter asks more broadly whether the researchers were interested in behaviors for which a ground truth could be defined. In other words, the Applicability criteria asks, regardless of how the researchers scored the responses, would it have been possible to score against a ground truth?  
For the second criteria, we analyzed whether the problem put to participants was one which a rational agent could best respond to in principle, regardless of whether the authors analyzed the results assuming the same problem definition.
For the third criteria, we considered flat scoring rules (i.e., paying every participant the same amount regardless of response) unambiguous, since they clearly assign the same score to every response. In cases where other rules were verbally described only (e.g., "Do your best to submit accurate responses") but a flat rule was used, we considered the flat reward to be the scoring rule for incentives for that problem since it is the rule that a rational agent would respond to.

In coding the fourth item--Was a flat scoring rule used to incentivize participants?--we encountered some cases where the researchers paid participants a given amount regardless of quality but filtered responses in analysis to only those that correctly answered one or more ``check'' questions. We coded such cases as using a flat rule unless researchers specifically noted that they did not pay participants who failed these questions. 

\pagebreak
\subsection{Coding results}
\label{coding_table}

\begin{table}[htb]
\begin{tabular}{llllll}
Study ID & Applicable? & Evaluative? & Well defined?   & Same rule?      & Flat rule?          \\
1        & Y           & Y           & N               & N               & Y                   \\
2        & Y           & Y           & Y               & Y               & N                   \\
3        & Y           & Y           & Y               & N               & N                   \\
4        & Y           & Y           & Y               & Y               & N                   \\
5        & Y           & Y           & Y               & Y               & N                   \\
6        & N           & NA          & NA              & NA              & NA                  \\
7        & Y           & Y           & N               & N               & Y                   \\
8        & Y           & Y           & N               & N               & Y                   \\
9        & Y           & Y           & N               & N               & Y                   \\
10       & Y           & Y           & N               & N               & NA                  \\
11       & Y           & Y           & N               & Y               & N                   \\
12       & Y           & Y           & Not enough info & N               & N                   \\
13       & Y           & Y           & N               & N               & N                   \\
14       & Y           & Y           & N               & N               & Y                   \\
15       & Y           & Y           & Not enough info & N               & Y                   \\
16       & Y           & Y           & N               & N               & N                   \\
17       & N           & NA          & NA              & NA              & NA                  \\
18       & N           & NA          & NA              & NA              & NA                  \\
19       & Y           & Y           & N               & Not enough info & Not enough info     \\
20       & Y           & Y           & Y               & N               & N                   \\
21       & Y           & Y           & Y               & N               & N                   \\
22       & Y           & Y           & N               & Not enough info & Not enough info     \\
23       & Y           & Y           & Y               & Y               & Y                   \\
24       & Y           & Y           & Not enough info & N               & Y                   \\
25       & Y           & Y           & N               & N               & Not enough info     \\
26       & N           & NA          & NA              & NA              & NA                  \\
27       & Y           & Y           & N               & Not enough info & Not enough info     \\
28       & Y           & Y           & N               & N               & Y                   \\
29       & Y           & Y           & Partially       & N               & N                   \\
30       & Y           & Y           & Not enough info & N               & N                   \\
31       & Y           & Y           & N               & Sometimes       & Sometimes           \\
32       & Partially   & Y           & N               & Not enough info & Not enough info     \\
33       & Y           & Y           & Partially       & N               & N                   \\
34       & Y           & Y           & N               & Y               & N                   \\
35       & Y           & Y           & N               & N               & Y                   \\
36       & Y           & Y           & N               & N               & Y                   \\
37       & Y           & Y           & Partially       & N               & Y                   \\
38       & N           & NA          & NA              & NA              & NA                  \\
39       & N           & NA          & NA              & NA              & NA                  \\
40       & N           & NA          & NA              & NA              & NA                  \\
41       & Y           & Y           & N               & Not enough info & Not enough info     \\
42       & Y           & Y           & N               & Not enough info & Not enough info     \\
43       & Y           & Y           & N               & N               & Y                   \\
44       & Y           & Y           & N               & N               & N                   \\
45       & Y           & Y           & N               & N               & Y (most conditions) \\
46       & Y           & Y           & N               & Y               & N                  
\end{tabular}
\end{table}

\end{document}